# User Interface Design for E-Learning Software

Behnam Faghih
Sama Technical and Vocational Training College
Islamic Azad University,
Boushehr Branch,
Boushehr, Iran
Behnam.Faghih@GMail.com

Dr. Mohammad Reza Azadehfar
Faculty of Music
University of the Arts
Tehran, Iran
Azadehfar@art.ac.ir

Prof. S. D. Katebi
Faculty of computer
Shiraz University, School of Electrical and Computer Engineering
Shiraz, Iran
Katebi@shirazu.ac.ir

*Abstract*—User interface (UI) is point of interaction between user and computer software. The success and failure of a software application depends on User Interface Design (UID). Possibility of using a software, easily using and learning are issues influenced by UID. The UI is significant in designing of educational software (e-Learning). Principles and concepts of learning should be considered in addition to UID principles in UID for e-learning. In this regard, to specify the logical relationship between education, learning, UID and multimedia at first we readdress the issues raised in previous studies. It is followed by examining the principle concepts of e-learning and UID. Then, we will see how UID contributes to e-learning through the educational software built by authors. Also we show the way of using UI to improve learning and motivating the learners and to improve the time efficiency of using e-learning software.

## I. INTRODUCTION

Teaching and learning have been changed over last twenty years, because of the ICT influences. The most important outcome of such changes is the ever-expansion of electronic education. E-learning devices should be developed based on the psychology of learners. Many studies have been done on the nature of learning and the factors that affect it. It is obvious that every design for e-learning should consider such studies in depth. Moreover, the principle of conventional education (non electronic) issues also have to be taken in board in e-learning by some adaptations. Even though, such adaptations raise some new problems which may conflict with the nature of distance learning from one way or another [1] [2].

One of the psychological matters that should be considered is User Interface (UI) in e-learning [1], because UI is the point of interaction between user and educational body. Aims of education may not be achieved if such correlation becomes unsuccessful, even if the educational content was selected well and the user is willing to learn [1] [3]. For this reason, main issues in successful correlations, which should be considered in User Interface Design for E-Learning (UIDEL), are in the focal point of this study.

## II. RELATED WORKS

Several works have taken place about the process of human learning and the factors that affect it [4, 6] and UI design [33]. However, there are very few works on UIDEL. One of the well-presented works on UIDEL is a book by Clark and Mayer [7] which addresses various studies on UIDEL, nonetheless, they looked at previous studied from the perspective of multimedia only. Moreover, in recent years, several studies have done about website designing [8] which in one way or another relates to concept of present study. After providing a brief introduction to literature of UID, multimedia, software engineer, and web design, we look at principles and concepts of UI followed by studying several practical suggestions for designing e-learning user interface.

### III. E-LEARNING

Teaching and learning through electronic devices, such as: computers, the Internet, web, TV, disks, telephone etc. is called e-learning[9]. However, this paper defines a specific description of e-learning: "teaching and learning using computer devices, memories and computer networks with the following characteristics:"

- All contents should be related to the educational aim;
- Curriculum must be designed according to the characteristics and tools of e-learning;
- The factors affecting education must be considered;
- Combining individual and corporation learning and group teaching must benefit from aappropriate method;
- Using multimedia such as: voice, picture, text etc. have a crucial role in presentation of the lessons.

Researches show[10]that humans learn:
- 10% of the things that they read
- 20% of the things that they hear
- 30% of the things that they see
- 50% of the things that they see and hear together
- 70% of the things that speak with others
- 80% of the things that they experience
- 95% of the things that they teach to other people

Also it should be noted that: about 75% of the general human learning is by the use of sense of sight, 13% by sense of hearing, 6% by sense of touch, 3% by sense of smell, and 3% by sense of taste[11].

E-learning makes it easy to learn lessons by seeing, hearing, discussion, experience and teaching to others. In fact obtaining all of the senses in forms of conventional learning (without using new technologies) is far difficult to reach.

Materials and equipment of training in the from of e-learning also can have a supportive and simplifying role in teaching-learning process, such as[4][12]:

- Facility of effective relation between learner and lessons using sight and hearing:
- Continuation of keeping student interested in subject.

In e-learning, various disciplines are involved.In fact, to have a successful results one should consider psychological, political, social, scientific, educational, economic, behavioral and technical issues at the same time. Over all issues involved, in this paper we only focus on psychological issues of user interface design.

### IV. USER INTERFACE (UI)

User interface is an interactive communication between user and computer software codes. Indeed, UI designer arranges elements (such as: multimedia, and tools like: Textbox, Label etc.) with which user can use computer more easily [13].

If software is difficult to use, if it forces user into mistakes, or if it frustrates user's efforts to accomplish his/her goals, he or shewill dislike it, regardless of the computational power it exhibits or the functionality it offers; because it molds a user's perception of the software, the interface has to be right[13].

Design should begin with an understanding of the intended users, including profiles of their age, sex, physical abilities, education, cultural or ethnic background, motivation, goals and personality[6]. Thus, a UI may not be useful for all computer users, while it may be just useful to specific users.

There are three golden rules to design UI[5]: (1) Place the user in control; (2) Reduce the user's memory load; (3) Make the interface consistent.

*A. The user interface design process*

The user interface design process encompasses four distinct framework activities[13]: (1) User, task, and environment analysis and modeling; (2) Interface design;(3) Interface construction; (4) Interface validation.

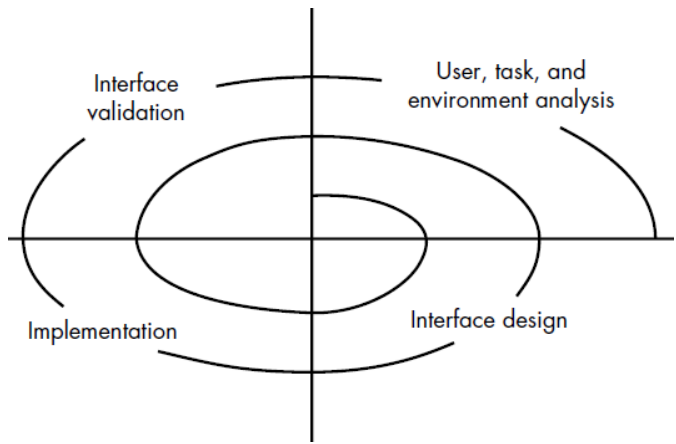

Figure 1 - The user interface design process

According to figure 1, the process of UID has a periodic nature in software's life cycle. Because users' interest, their skill level, new technologies, seeking diversity etc. keepsthe users to have ever increasing demand for changes.

*B. UI Design Issues*

Four common design issues almost always surface: System response time, help facilities available to user, error information handling, and command labeling. It is far better to establish each as a design issue to be considered at the beginning of software design, when changes are easy and costs are low [13].

## V. UI DESIGN ISSUES FOR E-LEARNING SOFTWARE

Knowledge about the factors that affect teaching and learning will enable teacher (e-learning designer)to make informed decisions in the teaching activities. This section analyses some important factors affecting teaching and learning, which are related to the UI. Also it presents a way to apply these factors in UI.

*A. Stimulus*

Learners need to be diligent and active since the main goal of education is improving their ideas, helping them come up with new notions, brushing up on their skills, habits and inclinations. Motivation is what causes us to act. Pedagogic stimulus should not involve fear of punishment or desire for reward, but it should involve willingness, the deep desire to learn something [4].

Pedagogic stimuli can shape up as[4]: (1)requirement, (2) motivation or (3) Coercion. To get the learning process started, at least one of the stimuli mentioned is needed.

**Applying stimulus in e-learning**

E-learning systems should be designed in such a way that the pervasive feeling of requirement and motivation grows constantly and the coercion feeling reduces. Some software application could have collaboration to each other. For example in [5] describes different requirements for collaboration between software application at architectural design level.

The following issues should be considered to increase motivation [4] [12] [7]:

- Curriculum
- How to arrange the layout and sequence of teaching materials
- Considering applications, benefits, and training objectives
- Tools
- Appearance and presentation of educational contents

Irrelevance of coercion on the education system leads to coercion exerting less effect on the education. However, by highlighting the requirements and motivation one can hope the learning process still goes on. Some suggestions to increase motivation in e-learning systems are as followed:

*1) Using speech interface*

Voice interface can help one overcome situations that are even highly stressful; it enables people to be more relaxed and creative. Moreover, Matching gender, personality, or accent fosters feelings of trust and interest [14].

A particular voice type can affect how much students like the speaker and how hard they try to understand the presented material[15].

*2) Using informal communication style instead of formal*

In order to stop computer looking like a piece of iron, Speech and text should be presented in an informal style to comfort the learner .Researches[16] show that under appropriate circumstances, people behave like a real person towardsa computer. Thus, to reach our purpose an informal style is more suitable. Nonetheless, the method must be approached carefully so as not to leave learners under the impression that the context is of no importance.Wordusage should not appear to be unreal. It is suggested to use the first person and second person pronouns instead of third-person pronouns.

Researches show that when people feel they are engaged in interactive conversation with the others, they try to understand more than when they just receive information[17].

Researches [18] demonstrate that students, who learn in an informal style, act better than the contrary.

It should be noted that conversation must be polite. Politeness of speech and behavior acts as a positive impact on learning[19].Furthermore, using polite words rather than imperative phrases increases learning [20].

Learners also gave more positive ratings to the on-screen agent who spoke with a human voice rather than a machine voice on an instrument designed to capture the social characteristics of a speaker[15].

*3) Using Animated Pedagogical Agent (APA)*

Educational agents are the shapes which appear in educational materials to suggest tips. These shapes can be a cartoon character, a video of a speaking head or specific virtual images (Avatars). Texts, speeches, illustrations, and animations can be used to form an agent [7].

The APA interface face-to-face creates an affective link with user/student[21].

One of the important concerns for an APA presentation is how to achieve the best speech and gesture cooperation [1].

Designers of multimedia learning environments should create life-like on-screen agents that speak in human voice rather than a machine-synthesized voice[15].

Researchers have proven that there is no difference between real and cartoon characters in training. Furthermore,a picture of the agent is not required to be shown while being used; only the voice would be sufficient[22].

*4) Using variety of colors in educational Medias*

Color variety, causes motivation and attention in students and makes educational materials seem more realistic [23].

Colors can also be used to group related information together. However, it should be noted that if more than 10 colors are involved, distortion of relation between information will be inevitable. Proper use of colors also accelerates reviewing and makes it easier. In addition, enough white space should be used on the interface screen. These spaces can bring order to pages to prevent clutter of information on screen. The last thing to be acknowledged about colors is to use dark text on a light background which expedites the reading[8].

*5) Learners having control over learning environment*

The learner control strategy has become more appreciated than tutor control or program control. In order to achieve this purpose,the following should be noted: Currently, the commonly used learner control strategies concentrate on the selection of topics, sequencing and pacing of contents, selection of instructional components, and control over learner-tutor and learner-learner interaction[2].

*6) Using background music*

Background music can help education become easier, more successful and more enjoyable. Other benefits are as follows[24]:

- Affects mood state
- Alters perception of time and space
- Affects physiological change
- Reduces stress and anxiety
- Enhances relaxation
- Causes arousal
- Motivates
- Be associated with product
- Enhances message reception
- Reduces noise distraction
- Aids concentration
- Aids memorization
- Increases task performance
- Enhances creativity
- Increases the enjoyment of mental and physical activity

When choosing the background music the following should be considered[24]:

   *a) Listening to music carries with it a cognitive processing load*

   *b) The diversity of different music styles cause different psychological effects*

   *c) Music is suitable or unsuitable to use as BM depending on its constituents*

Tempo, tonality and volume are the three constituents which require the most scrutiny

Simple musical structure is the best choice and also music should be light with average volume[25].

Simple structure of the music depends on the[24]: Rhythm, orchestration, harmony, form, etc. Recognition of this kind of music should be done by the professionals. It is recommended that background music won't be used if the designer is not a musical expertise, and in doubt whether to use it or not. The reason will be discussed.

*B. Role of working and long-time memory in learning*

Working memory is the mental activity center, where all active thoughts take place. Although this is a powerful processor, its capacity is limited. In fact new information must be processed in working memory, then merge with information into long-term memory, and at the end, when needed; it transfers from long-term memory to working memory to be used [7].

Teaching styles in electronic courses should help learners by using words and images in the course so that it combines with knowledge in long-term memory. To accomplish this, the following should be considered:

*1) Highlighting important contents in the lesson*

Techniques such as flash, changing colors, and big or/and bold text etc. should be used to highlight important texts and visual information. Another way to attract learner's attention to important contents is to write a list of learning objectives[7].

*2) Desaturation of working memory with unnecessary materials*

Working memory should be free to process new information. When a limited capacity of working memory is saturated, the process doesn't work properly and learning process becomes slow. Even keeping a limited amount of information and processing it at the same time is very difficult for the working memory[7].

For this purpose in electronic courses, irrelevant visuals should be minimized, background music and noisy sounds should be removed and also clear, concise and appropriate text should be used. However, as mentioned, background music under certain conditions can increase the efficiency of training. Thus, eliminating irrelevant information leads to mental pressure reductionwhich makes more capacity available to process information for the learning process[7].

Some e-learning designers use stories, background music, animations etc. for entertainment and motivation, which should be avoided because of the following reasons[7]:

- Disrupting Focus
- Causing pause
- Misleading information – integrating irrelevant and inappropriate information

Music and sounds are not suitable when learners are under increased mental pressure as they may overload working memory.Such as when educational contents are unfamiliar, contents are presented at high speed, or training speed is not controlled by the learner.

Researches also evidence that the additional words should be avoided[26][27].

However, this recommendation does not mean that adding interesting images under all circumstances is wrong, but they are problematic when they interfere with the efforts of the students trying to understand the course.

*3) Integration of auditory and visual information in working memory with existing knowledge in long-term memory*

Working memory integrates words with images to provide a unified structure, and then these ideas are combined with the existing knowledge in the long-term memory. This issue needs to be processed in the working memory. The textual and visual information being presented together, and not separately, makes it easier to integrate words with images [7][28].

Texts can be fragmented and placed in different parts of the image. When the text describes an operation or mode on the picture, it can be used as a pop-up message[7].

*4) Using elements of the real world*

Merely adding new knowledge into the long-term memory is not enough. New knowledge must be encrypted in their long-term memory so that when needed it is simple and quick to recover. Electronic lessons should offer objective and practical activities in the examples and exercises; that way long-term memory is a good position for new knowledge to be retrieved. The shape and appearance of the educational environment should be based upon the actual environment where the learners at the end of the course should present their acquired information[15].

*C. Using multimedia to present lessons*

It is recommended to use instructional materials through a combination of visual (such as texts, graphics, etc.) and auditory elements (such as speech, music, etc.). This suggestion is based on two principles[7]:

*a) Learning theory of mind;* this is based on these assumptions: a) Humans have separate channels for processing verbal and visual materials, b) Each of these channels can process a limited amount of information at any given time c) Students seriously try to build visual and verbal models by the presented materials and the relation between them.

*b) People understand the materials better, when they are fully engaged in the process.* Presenting a content using words and images make learners' mind more active, learners try to establish a relationship between the words and the images. Using the words solely can encourage students (especially learners who are less experienced) to learn superficially and learners would not be able to communicate between their prior knowledge and new contents.

According to the above principles, the following is recommended:

*1) Using speech instead of typing text to explain the graphic images (such as: plans, photographs, chart, animation etc.) on the screen*

This is due to the reason that there are two separate channels of information (visual and auditory) going to the brain which helps reduce the compression of information (Figure 2). It should be indicated that in the presence of a complex material such as a mathematical formula, the text might be required to be displayed on the page.

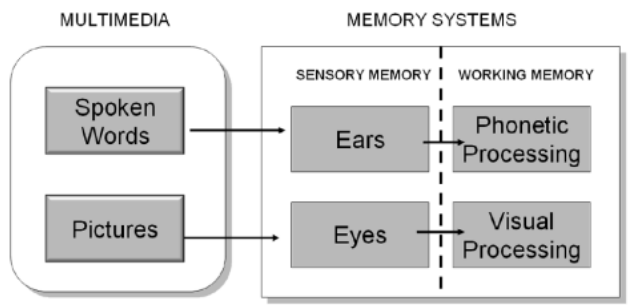

Figure 2 - Balancing content across visual and auditory channels with presentation of narration and graphics

Several studies have been conducted on this subject that approved the suggestion[29][30][22].Under the following conditions combination of text and speech can be (or should be) avoided[7]:

- There is no picture to present.
- Learner has enough time to process the images and words.
- Learner may have difficulty to process spoken words.
- When the learners need to see materials frequently (such as a guide for problem solving).

*D. Availability*

Availability means that users can easily access intended content. Owing to learners referring to previous taught material regularly or looking for specific contents,e-learning system should provide leaners with the ability to search the contents and help them attain their desired content easily. Moreover, whenever the words or phrases that are used in the text exist elsewhere, they should act as a link to navigate, describe, and return the user to the previous page simply[3][31]. Also cloud computing and in particular e-learning cloud template [32] is a new technology that allows IT department to provide high availability.

## VI. CONCLUSION AND DISCUSSION

E-learning is not merely making the traditional educational materials electronic. In e-learning UI plays a key role in achieving educational objectives. Psychological issues in learningare new findings in e-learning, which influence the design of UI.

As it was mentioned before, to provide stimulus, speech interface, informal style, agents, learner control environment, colors and background music must be used in presentation. The working memory also has a crucial role in learning and it should not be saturated with unnecessary information. When using multimedia information, it's better to use both hearing and vision channels if possible.

The UID should not only be seen as an artistic phenomenon, but artistic tools such as, graphics, music, animation, etc., must be compatible with the educational psychological issues.

Statistics and reports indicate that there are many dropouts in e-learning inclusive control[7]. It seems that one of the most important challenges is the UI of software. It is required to examine the relation between the UI and dropouts.

## VII. FUTURE WORKS

- Consider how UI should be designed for practical learning, such as: music, painting, carpentry etc. since they're different from theoretical subjects and learners need to learn knowledge of using tools.

- Consider e-learning environment, such as: examples, exercises, duration of lessons etc. The order of examples and exercises through the course should be specifically determined. Also, duration of each lesson should be analyzed, because how many minutes conventional classes last is based upon traditional educational environments. Thus, duration of e-learning classes must be based on e-learning environment.